\documentclass{ws-ijmpd}
\usepackage[super,compress]{cite}
\usepackage{color}
\usepackage{ulem}
\begin{document}

\markboth{Garc\'{\i}a-Aspeitia, Maga\~na, Hernandez-Almada, Motta}{}

%
\catchline{}{}{}{}{}
%

\title{Probing dark energy with braneworld cosmology in the light of recent cosmological data}

\author{Miguel A. Garc\'{\i}a-Aspeitia}

\address{Consejo Nacional de Ciencia y Tecnolog\'ia, \\ Av. Insurgentes Sur 1582. Colonia Cr\'edito Constructor, Del. Benito Ju\'arez C.P. 03940, Ciudad de M\'exico, M\'exico.\\
Unidad Acad\'emica de F\'isica, Universidad Aut\'onoma de Zacatecas, Calzada Solidaridad esquina con Paseo a la Bufa S/N C.P. 98060, Zacatecas, M\'exico. \\
aspeitia@fisica.uaz.edu.mx}

\author{Juan Maga\~na}

\address{Instituto de F\'isica y Astronom\'ia, Facultad de Ciencias, Universidad de Valpara\'iso, Avda. Gran Breta\~na 1111, Valpara\'iso, Chile. \\
juan.magana@uv.cl}

\author{A. Hern\'andez-Almada}

\address{Facultad de Ingenier\'ia, Universidad Aut\'onoma de Quer\'etaro, Centro Universitario Cerro de las Campanas, 76010, Santiago de Quer\'etaro, M\'exico.\\
ahalmada@uaq.mx\\
Instituto de F\'isica, Universidad Nacional Aut\'onoma de M\'exico, Circuito Exterior C.U., A.P. 20-364, Ciudad de M\'exico, 04510, M\'exico.}

\author{V. Motta}

\address{Instituto de F\'isica y Astronom\'ia, Facultad de Ciencias, Universidad de Valpara\'iso, Avda. Gran Breta\~na 1111, Valpara\'iso, Chile. \\
veronica.motta@uv.cl}

\maketitle

\begin{history}
\received{Day Month Year}
\revised{Day Month Year}
\end{history}

\begin{abstract}
We investigate a brane model based on Randall-Sundrum scenarios
with a generic dark energy component. The latter drives the accelerated expansion at late times of the Universe. In this scheme, extra terms are added into Einstein Field equations that are propagated to the Friedmann equations.
To constrain the dark energy equation of state (EoS) and the brane tension we use observational data with different energy levels (Supernovae type Ia, $H(z)$, baryon acoustic oscillations, and cosmic microwave background radiation  distance, and a joint analysis) in a background cosmology. 
Beside EoS being consistent with a cosmological constant 
at the $3\sigma$ confidence level for each dataset, the baryon acoustic oscillations probe favors an EoS consistent with a quintessence dark energy. Although we found different lower limit bounds on the brane tension for each data sets, being the most restricted for CMB, there is not enough evidence of modifications in the cosmological evolution of the Universe by the existence of an extra dimension within observational uncertainties.
Nevertheless, these new bounds are complementary to those obtained by other probes like table-top experiments, Big Bang Nucleosynthesis, and stellar dynamics. Our results show that a further test of the braneworld model with appropriate correction terms or a profound analysis with perturbations, may be needed to improve the constraints provided by the current data.
\end{abstract}


\ccode{PACS numbers: 98.80.Cq, 26.35.+c, 98.80.Ft, 98.70.Vc}


\section{Introduction} \label{Int}

Several cosmological observations of Supernovae of the Type Ia (SNIa) at high redshift,
show evidence of an accelerated expansion of the Universe at late times 
\cite{Schmidt,Perlmutter,Riess}.  
This is also supported by the observations of anisotropies in CMB and baryon acoustic oscillations (BAO).  
In the standard cosmological scenario, the responsible for the Universe accelerated expansion 
is an entity which made up the $\sim70\%$ of its total content and it is dubbed as dark energy \cite{PlanckCollaboration2013}.
Several models try to explain this late time cosmic trend \cite{Shi:2012ma,Magana:2014voa,Magana:2015wra}, but the most favored candidate by the
cosmological data is still the cosmological constant \cite{PlanckCollaboration2013}.  
However, the latter shows conceptual and theoretical problems when assuming that  
its energy density arises from quantum vacuum fluctuations \cite{Weinberg,Zeldovich}. 
Under this supposition, the theoretical prediction of the CC energy density differs
$\sim120$ orders of magnitude from the observational estimations \cite{Weinberg,Zeldovich}. 
The other well-known difficulty of the CC is the coincidence problem, i.e. why DE density is similar to that of the dark matter (DM) component today \cite{Weinberg}.

In this vein, the CC problems have motivated the appearance of alternative candidates for DE, being some of the most popular: 
the quintessence, phantom field, Chaplygin gas, Holographic DE, among others
(see \cite{Copeland:2006wr} for an excellent DE models review).
An interesting paradigm is to consider extra dimensions of space-time
which could be the source of the current accelerated expansion. 
For instance, the Dvali-Gabadadze-Porrati (DGP) model \cite{Dvali:2000hr,Deffayet:2001pu} 
generates a natural accelerated expansion with a geometrical threshold associated 
to a five-dimensional space-time. However, a DE component must be added 
to achieve a stable late cosmic acceleration.
Another example are the models
proposed by Randall and Sundrum (RSI or RSII) \cite{Randall-I,Randall-II}
that have the added benefit of providing a solution for the hierarchy and the CC quantum vacuum fluctuations problems
(see \cite{m2000,Martinez,GarciaAspeitia:2011xv} for more details).
Nevertheless, RS models drive a late time acceleration only by adding a dark energy field. 

In a covariant approach of the RSII models, the Einstein's field equations are modified 
assuming a five dimensional bulk with Schwarzschild-Anti'dSitter (S-$\rm AdS_{5}$) geometry and a four dimensional manifold embedded in this bulk, called the \emph{brane}. 
Note that for cosmological purposes the brane is considered as a FLRW structure, 
but in general it can take any geometry. 
The main modifications to the Einstein field equations 
lie in three new tensors: the first one considers second order corrections 
to the energy-momentum tensor; the second one allows matter in the bulk 
and the last one takes into account non-local effects associated with the Weyl's tensor \cite{sms}.
An important term in the theory is the brane tension, $\lambda$, which shows the threshold between the corrections that come from branes and those who belong to the traditional Einstein's equation. 
These brane correction terms could
produce important changes in the Universe  dynamics that can be tested using the latest astrophysical or cosmological observations.

Our main goal is to investigate the effect of one extra dimension 
on the background cosmology, mainly the DE properties. 
Indeed, we focus our inquiry on what is the preferred equation 
of state (EoS) for DE in brane models? and what is the constraint for the brane tension provided by current cosmological observations?
In the present model, DE is located in our brane, together with the other Universe components (baryons, radiation and dark matter), 
with the restriction that gravity is the only interaction that can overstep to the extra dimension. The condition for the dark energy EoS is that it must always fulfill a generalized inequality,  
shown in Ref. \cite{Maartens:2003tw}, to obtain an accelerated dynamic on the brane. 
As it happens in General Relativity (GR), DE is divided in Quintessence ($-1<\omega_{de}<-1/3$), CC ($\omega_{de}=-1$), and Phantom field ($\omega_{de}<-1$), but parametrized by the extra dimensions \cite{Copeland:2006wr}. 
As mentioned before, although the late cosmic acceleration does not emerge
from the extra dimension, the brane dynamics can help us to understand
the dark energy and cosmic acceleration in this kind of scenario \cite{m2000,Martinez,GarciaAspeitia:2011xv}. 

Recent results using RS frame have only assumed a geometrical point of 
view (i.e. no DE component, see \cite{Wang:2006ue,Alam:2016wpf} as interesting examples). 
However, a robust analysis of the DE dynamic in a simple brane scenario is needed. 
In this work, we test a RS-like model 
that has all the basic components of the Universe, including a DE with a generic EoS, to constrain its parameters using recent cosmological observations at different energy scales. This kind of test also allow us to confirm whether the model is consistent at high and low energies in the cosmological evolution (i.e. using observational probes at different redshifts).
To estimate the dark energy EoS 
and the brane tension, we use SNIa, $H(z)$, BAO and CMB distance constraints. These new bounds are complementary to those obtained by other probes like table-top (TT) experiments, 
Big Bang Nucleosynthesis (BBN), and stellar dynamics \cite{Gergely:2006xr}.

The paper is organized as follows: in Sec. \ref{EM} we show the modified Einstein's equation 
by the presence of branes in the RS scenario. We find the modified Friedman equation assuming matter, radiation and a generic DE as the Universe components; 
in addition, we get the deceleration parameter in terms of brane corrections. 
In Sec. \ref{data} we perform a statistical analysis to constrain the EoS of the dark energy
and the brane tension using various observations as $H(z)$ measurements, SNIa, BAO, CMB distance and finally a joint analysis.
In Sec. \ref{Res}, we present and discuss the results obtained with the analysis of the previous section and finally in Sec. \ref{CR} we give some conclusions and remarks.
In what follows, we work in units with $c=\hbar=1$, unless explicitly written.

\section{Brane cosmology} \label{EM}
First of all, we introduce the Einstein's field equation projected onto the brane
\begin{equation}
G_{\mu\nu}+\xi_{\mu\nu}=\kappa^2_{(4)}T_{\mu\nu} + \kappa^4_{(5)}\Pi_{\mu\nu} + \kappa^2_{(5)}F_{\mu\nu}, \label{1}
\end{equation}
where $T_{\mu\nu}$ is the four-dimensional energy-momentum tensor of the matter trapped in the brane, 
$G_{\mu\nu}$ is the classical Einstein's tensor and the rest of terms in the right and left sides 
of this equation are explicitly given by:
\begin{subequations}
\begin{eqnarray}
\kappa^2_{(4)}&=&8\pi G_{N}=\frac{\kappa^4_{(5)}}{6}\lambda, \\
\Pi_{\mu\nu}&=&-\frac{1}{4}T_{\mu\alpha}T_{\nu}^{\alpha}+\frac{TT_{\mu\nu}}{12}+\frac{g_{\mu\nu}}{24}(3T_{\alpha\beta}T^{\alpha\beta}-T^2), \\
F_{\mu\nu}&=&\frac{2T_{AB}g_{\mu}^{A}g_{\nu}^{B}}{3}+\frac{2g_{\mu\nu}}{3}\left(T_{AB}n^An^B-\frac{^{(5)}T}{4}\right), \\
\xi_{\mu\nu}&=&^{(5)}C^E_{AFB}n_En^Fg^{A}_{\mu}g^{B}_{\nu}.
\end{eqnarray}
\end{subequations}
Here $G_N$ is the Newton's gravitational constant, $\lambda$ is the previously mentioned brane tension, 
$\kappa_{(4)}$ and $\kappa_{(5)}$ are the four- and five-dimensional coupling constants of gravity respectively.
The tensor $\Pi_{\mu\nu}$ represents the quadratic corrections on the brane generated by the energy-momentum tensor, 
$F_{\mu\nu}$ gives the contributions of the energy-momentum tensor in the bulk, which is 
projected onto the brane through the unit normal vector $n_A$.
The tensor $\xi_{\mu\nu}$ provides the contribution of the five-dimensional Weyl's tensor
projected onto the brane manifold \cite{sms} \footnote{Notice that the latin letters take the values $0,1,2,3,4$.}. 
It is worth to note that non-local corrections are negligible in cosmological cases \cite{m2000}, under the assumption of a AdS$_{(5)}$ bulk. 

To derive the Friedmann equations under the modified field equations,
we consider an homogeneous and isotropic 
Universe in which a line element is given by
\begin{equation}
ds^2=-dt^2+a(t)^2(dr^2+r^2(d\theta^2+\sin^2\theta d\varphi^2)),
\label{ds}
\end{equation}
where $a(t)$ denotes the scale factor. 
The recent Planck measurements \cite{PlanckCollaboration2013} suggest
a curvature energy density $\Omega_k\simeq 0$, thus we assume a flat geometry.
We consider radiation and dark matter components as perfect fluids in the brane. We assume that the bulk has no matter component.
By combining Eqs. \eqref{1} to \eqref{ds}, we obtain the modified Friedmann equation:
\begin{equation}
H^2=\kappa^2\rho_{eff}, \label{FLRWM}
\end{equation}
where
\begin{equation}
\rho_{eff}=\sum_i\rho_i\left(1+\frac{ \rho_i }{2\lambda} \right),
\end{equation}
defining $\kappa^2=8\pi G/3=\kappa^2_{(4)}/3$ as the \emph{renamed} gravitational coupling constant.
$H=\dot{a}/a$ is the Hubble parameter, and $\rho_i$ is the energy density for the radiation, dark matter and DE.
It is worth to notice that the low energy regime, i.e. the canonical Friedmann equation, is recovered when $\rho_i/2\lambda\to0$. Crossed terms were not used in the Friedmann equation, i.e. there is not interaction between different species. In addition, if we consider, for instance, that the bulk black hole mass vanishes, the bulk geometry reduces to $\rm AdS_{5}$ and $\rho_{\epsilon }=0$ \cite{m2000,MaartensCos}.

As a complement, we write the modified Friedmann equation at high energy regime as:
\begin{equation}
H^2_{high}=\kappa^2\sum_i\frac{ \rho_i^2}{2\lambda}, \label{FLRWM2}
\end{equation}
where it is assumed that $\rho_i/2\lambda\gg1$. The latter is the high energy limit because we are assuming that the mean density of the fluids is much higher than the brane tension, so that any brane corrections in the equations of motion are highly suppressed by the brane energy scale. This limit is specially used in inflationary cosmology where the effects are more noticeable (for an excellent review of brane world inflation see \cite{Lidsey:2003ws}).

As mentioned before, we assume that the EoS of DE satisfies the constraint 
\begin{equation}
\omega_{de}<-\frac{1}{3}\left[\frac{1+2\rho_{de}/\lambda}{1+\rho_{de}/\lambda}\right], \label{omega}
\end{equation}
to obtain an accelerated Universe and this value must be constrained via the cosmological data. Thus, the Friedmann equation can be written as:

\begin{eqnarray}
H^2=\kappa^2\left[\frac{\rho_{0m}}{a^3}\left(1+\frac{\rho_{0m}}{2\lambda a^3}\right)+
\frac{\rho_{0r}}{a^4}\left(1+\frac{\rho_{0r}}{2\lambda a^4}\right)
+\frac{\rho_{0de}}{a^{3(1+\omega_{de})}}\left(1+\frac{\rho_{0de} }{2\lambda a^{3(1+\omega_{de})}}\right) \right]. \label{eq1}
\end{eqnarray}
Using the density parameters, $\Omega_{i} \equiv \rho_{i}/\rho_{crit}$, and 
redshift, Eq. \eqref{omega} and \eqref{eq1} can 
be written as:
\begin{eqnarray}
&&E(z)^{2}=\Omega_{0m}(1+z)^{3}+
\Omega_{0r}(1+z)^{4}+\Omega_{0de}(1+z)^{3(1+\omega_{de})}\nonumber\\
&&+\mathcal{M}\left[\Omega_{0m}^2(1+z)^{6}+\Omega_{0r}^2(1+z)^{8}+\Omega_{0de}^2(1+z)^{6(1+\omega_{de})}\right], \qquad
\label{eq:H}
\end{eqnarray}
where $E(z)^2\equiv H(z)^2/H_0^2$, $\Omega_{r}=2.469\times10^{-5}h^{-2}(1+0.2271 N_{eff})$, $N_{eff}=3.04$ is
the standard number of relativistic species \cite{Komatsu:2011} and
\begin{equation}
\mathcal{M}\equiv \frac{H_0^2}{2\kappa^2\lambda}=\frac{\rho_{crit}}{2\lambda}, \label{const}
\end{equation}
being $H_0$ the Hubble constant, and $\rho_{crit}$ the Universe critical density. 
Notice that when $\mathcal{M}\to0$, the canonical 
Friedmann equation with $w_{de}$ is recovered. 
If $w_{de}=\omega_{\Lambda}=-1$, i.e. the DE is the CC, 
we obtain the traditional $\Lambda$CDM dynamics.

At early times the brane dynamics dominate over other terms in the Universe,
buy is negligible at late time. Indeed, given a value for the brane tension, we can infer 
the limits of high and low energies in terms of the redshift:
$z+1\gg\sum_i(\lambda/\rho_{0i})^{1/3(1+\omega_i)}$ and
$z+1\ll\sum_i(\lambda/\rho_{0i})^{1/3(1+\omega_i)}$ respectively. For example, in matter domination epoch, the previous expressions can be rewritten as:
$z\gg(\lambda/\rho_{0m})^{1/3}-1$ and $z\ll(\lambda/\rho_{0m})^{1/3}-1$,
for high and low energy limits respectively.

In addition, from Eq. \eqref{omega} the DE EoS should satisfy the following constraint to obtain a late cosmic acceleration:
\begin{equation}
\omega_{de}<-\frac{1}{3}\left[\frac{1+4\mathcal{M}\Omega_{0de}(1+z)^{3(1+\omega_{de})}}{1+2\mathcal{M}\Omega_{0de}(1+z)^{3(1+\omega_{de})}}\right],
\label{eq:de_constraint}
\end{equation}

On the other hand, using Eq. (\ref{eq:H}) the deceleration parameter, $q(t)\equiv-\ddot{a}(t)/a(t)H(t)^2$,
can be written in terms of redshift as

\begin{equation}
q(z)=\frac{q_{I}(z)+\mathcal{M}\,q_{II}(z)}{E(z)^2},
\label{eq:qz}
\end{equation}
where
\begin{eqnarray}
q_{I}(z)=\frac{\Omega_{0m}}{2}(1+z)^3+\Omega_{0r}(1+z)^4+
\frac{\Omega_{0de}}{2}(1+3\omega_{de})(1+z)^{3(1+\omega_{de})},\nonumber\\
q_{II}(z)=2\Omega_{0m}^2(1+z)^6+3\Omega_{0r}^2(1+z)^8+
\Omega_{0de}^2(2+3\omega_{de})(1+z)^{6(1+\omega_{de})}.
\end{eqnarray}
In the same way, we recover the traditional behavior for $q(z)$ when $\mathcal{M}\to0$, i.e. when brane effects are negligible.

\section{Data and Methodology} \label{data}

In order to constrain the EoS of the dark energy within the braneworld geometry dynamics
we will perform a Markov Chain Monte Carlo (MCMC) analysis using the recent cosmological data of the Hubble parameter measurements, 
SNIa, BAO, and CMB from Planck data release 2015. We assume a Gaussian likelihood $\mathcal{L}\propto \exp(-\chi^{2}/2)$, 
which depends on the model parameters. We consider Gaussian priors on $h$, and $\Omega_{b}h^{2}$  
(see Table \ref{tab:priors}), and the only free parameters of the analysis are $\Omega_{m}$, $\omega_{de}$, and $\mathcal{M}$.
The affine-invariant MCMC method implemented in \emph{emcee} package \cite{emcee:2013} is used 
to find the confidence region of these free parameters.
In the following we present the data and how the merit functions, $\chi$, are constructed.

\begin{table}
\tbl{Priors on the different parameters of the brane-world model. For
$h$ and $\Omega_b h^{2}$ we use the values given by \cite{Riess:2016} and \cite{Cooke:2014} respectively.}
{\begin{tabular}{@{}cc@{}} \toprule
Parameter&Allowance\\ \colrule
$h$& $0.7324\pm{0.0174}$ (Gaussian) \\
$\Omega_b h^{2}$ & $0.02202\pm 0.00046$ (Gaussian)\\
$\Omega_{m}$& $[0,1]$ (Uniform)\\
$w_{de}$& $[-2.5,0]$ (Uniform)\\
$\mathcal{M}$&$[0,0.5]$ (Uniform)\\ \botrule
\end{tabular} \label{tab:priors}}
\end{table}

\subsection{$H(z)$ measurements}\label{subsec:hz}

The measurements of the expansion rate of the Universe as a function of redshift,
i.e. $H(z)$, are widely used to test cosmological models.
They are the most direct and model independent observables of the dynamics of the Universe.
Thus, any evidence of an extra dimension should be reflected in the fitting of these data. 
Nevertheless, the $H(z)$ data rely on a low-redshift range and since that the brane corrections are only important at high energy regime,
we expect to obtain low-significance brane parameters.
We use $34$ data points compiled by \cite{Sharov:2014voa} which span the redshift range $0.07< z <2.3$.

We start by writing the merit function, $\chi^2_{H}$, as:
\begin{equation}
\chi_{H}^2 = \sum_{i=1}^{34} \frac{ \left[ H_{th}(z_{i})-H_{obs}(z_{i})\right]^2 }{ \sigma_{H_i}^{2} },
\end{equation}
where $H_{obs}(z_{i})$ is the observed Hubble parameter at $z_{i}$,
$\sigma_{H_i}$ is the error function, and $H_{th}(z_{i})$ the theoretical value given by Eq. (\ref{eq:H}).
It is worth to notice that some $H(z)$ points have been derived from the BAO information.
Through this work we assume there is no correlation between $H(z)$ and BAO.

\subsection{Type Ia Supernovae (SNIa)}\label{subsec:snia}
The observations of distant SNIa is the classical test to
probe the late cosmic acceleration and the nature of dark energy. Although these observations
 could be also used to test modified gravity, \cite{BraneSN:2006} found that the SNIa data provide
a low-significance in the brane model parameter estimation. To validate this method to constrain brane parameters,
we use the Lick Observatory Supernova Search (LOSS) sample containing $586$ SNIa 
in the range $0.01<z<1.4$ \cite{Ganeshalingam:2013mia}. 
The relation between the distance modulus $\mu$ and the luminosity distance $d_L$ is given by
\begin{equation}
\mu(z) = 5\log_{10} [d_L(z)/{\rm Mpc}] + \mu_0,
\end{equation}
where $\mu_0$ is a nuisance parameter and 
\begin{equation}
d_L(z) = (1+z)\int_0^z \frac{dz'}{E(z')}\,.
\end{equation}
The expression for $E(z)$ was presented in Eq. \eqref{eq:H}. After we marginalize over $\mu_0$, 
the SNIa constraints are obtained by minimizing the function $\chi^2_{\rm SNIa} = A - B^2/C$, where
\begin{eqnarray}
A &=& \sum_{i=1}^{586} \frac{[\mu(z_i) - \mu_{{\rm obs}}]^2}{\sigma_{\mu_i}^2}, \nonumber \\
B &=& \sum_{i=1}^{586} \frac{\mu(z_i) - \mu_{{\rm obs}}}{\sigma_{\mu_i}^2},  \\
C &=& \sum_{i=1}^{586} \frac{1}{\sigma_{\mu_i}^2}\,. \nonumber
\end{eqnarray}

\subsection{Baryon acoustic oscillations} \label{subsec:bao}

Baryon acoustic oscillations are the signature of the interactions of 
baryons and photons in a hot plasma on the matter power spectrum in the pre-recombination epoch. 
Since BAO measurements are standard rulers, they are used as a geometrical probe to constrain cosmological 
parameters of modified gravity and dark energy models.
The different surveys usually give the BAO information in the ratio $D_{V}/r_{s}(z_{d})$, where
the distance scale $D_V$ is defined as
\begin{equation}
D_V(z)=\frac{1}{H_0}\left[(1+z)^2d_A(z)^2\frac{z}{E(z)}\right]^{1/3},
\end{equation}
$d_A(z)$ is the Hubble-free angular diameter distance
which relates to the Hubble-free luminosity distance through $d_A(z)=d_L(z)/(1+z)^2$. 
This scale is calibrated using the comoving sound horizon radius, $r_{s}$,
at the end of the drag epoch (typically derived from CMB) defined as

\begin{table}
\tbl{BAO data from different surveys: six-degree-Field Galaxy Survey (6dFGS), WiggleZ experiment, 
Sloan Digital Sky Survey (SDSS) Data Release 7 (DR7), Baryon Oscillation Spectroscopic Survey
(BOSS)-SDSS DR9, and BOSS-SDSS DR11.}
{\begin{tabular}{@{}cccc@{}} \toprule
Quantity& $z$& BAO measurement& Survey\\ \colrule
$d_{z}\equiv\frac{r_s(z_d)}{D_V(z)}$ & $0.106$ & $0.336\pm0.015$& 6dFGS \cite{Beutler2011:6dF}\\
$d_{z}$ & $0.44$& $0.0870\pm0.0042$  & WiggleZ \cite{Kazin:2014,Gong:2015}\\
$d_{z}$ & $0.6$& $0.0672\pm0.0031$   & WiggleZ \cite{Kazin:2014,Gong:2015}\\
$d_{z}$ & $0.73$& $0.0593\pm0.0020$  & WiggleZ \cite{Kazin:2014,Gong:2015}\\
$d_{z}$ & $0.15$ & $0.2239\pm0.0084$ & SDSS DR7 \cite{Ross:2014}\\
$d_{z}$ & $0.32$ & $0.1181\pm0.0022$ & SDSS-III BOSS DR11\cite{Anderson:2014}\\
$d_{z}$ & $0.57$ & $0.0726\pm0.0007$ & SDSS-III BOSS DR11\cite{Anderson:2014}\\
$\frac{D_{H}(z)}{r_s(z_d)}$& $2.34$  & $9.18\pm0.28$ & SDSS-III BOSS DR11 \cite{Delubac:2014aqe}\\
$\frac{D_{H}(z)}{r_s(z_d)}$& $2.36$  & $9.00\pm0.3$ & SDSS-III BOSS DR11 \cite{Font-Ribera:2014}\\ \botrule
\end{tabular} \label{tab:bao}}
\end{table}

\begin{equation}
r_s(z) =  \int_z^\infty \frac{c_s(z')}{H(z')}dz',
\end{equation}
where the sound speed $c_s(z) = 1/\sqrt{3\left(1+\bar{R_b}/\left(1+z\right)\right)}$, with
$\bar{R_b} = 31500\, \Omega_{b}h^2(T_{CMB}/2.7\rm{K})^{-4}$, and
$T_{CMB}$ is the CMB temperature. 
The redshift $z_d$ at the baryon drag epoch is fitted with the
formula proposed by \cite{Eisenstein:1997ik},
\begin{equation}
z_d =\frac{1291(\Omega_{m}h^2)^{0.251}}{1+0.659\,(\Omega_{m}h^2)^{0.828}}\left[1+b_1\left(\Omega_bh^2\right)^{b_2}\right],
\label{eq:zd}
\end{equation}
where
\begin{eqnarray}
b_1 &=& 0.313\left(\Omega_{m}\,h^2\right)^{-0.419}\left[1+0.607\left(\Omega_{m}\,h^2\right)^{0.674}\right], \\
b_2 &=& 0.238\left(\Omega_{m}\,h^2\right)^{0.223}.
\label{eq:bzd}
\end{eqnarray}
It is worth to note that $r_{s}(z_{d})$ depends on the early time physics, where the brane corrections could be important due to the term $\rho/\lambda$,
and that the Eqs. (\ref{eq:zd}-\ref{eq:bzd}) were derived for the standard cosmology neglecting the possible brane effects. 
We assume as first approximation, that they are valid for our brane model. Although the BAO data provide stronger brane constraints, 
they could be biased due to the standard model assumption (actually, a $\mathcal{M}$ bound consistent with $\Lambda$CDM is expected). 
Therefore, a perturbation theory for this brane model is needed to estimate straightforward constraints. 

In our analysis, we use the BAO data shown in Table \ref{tab:bao}.

Notice that the high redshift points are given in terms of $D_{H}(z)=H(z)^{-1}$. The function-of-merit for all the BAO data points, $\chi^2_{BAO}$ is:

\begin{equation}
\chi^2_{BAO} = \chi^2_{6dFGS}+\chi^2_{WiggleZ} +
\chi^2_{DR7} + \chi^2_{DR11A} + \chi^2_{DR11B},
\end{equation}
where
\begin{eqnarray}
&&\chi^2_{6dFGS} =\left(\frac{d_z(0.106)-0.336}{0.015}\right)^2,\nonumber\\
&&\chi^2_{WiggleZ} = \left(\frac{d_z(0.44)-0.0870}{0.0042}\right)^2+\left(\frac{d_z(0.6)-0.0672}{0.0031}\right)^2\nonumber\\&&+\left(\frac{d_z(0.73)-0.0593}{0.0020}\right)^2,\nonumber\\
&&\chi^2_{DR7}=\left(\frac{d_z(0.15)-0.2239}{0.0084}\right)^2,\nonumber\\
&&\chi^2_{DR11A}=\left(\frac{d_z(0.32)-0.1181}{0.0023}\right)^2+\left(\frac{d_z(0.57)-0.0726}{0.0007}\right)^2,\nonumber\\
&&\chi^2_{DR11B}=\left(\frac{\frac{D_H(2.34)}{r_{s}(z_{d})}-9.18}{0.28}\right)^2+\left(\frac{\frac{D_H(2.36)}{r_{s}(z_{d})}-9.00}{0.3}\right)^2.
\end{eqnarray}
  
\subsection{CMB distance constraints from Planck 2015 measurements} \label{subsec:cmb}

The anisotropy measurements in the temperature of CMB radiation provide narrow constraints
on cosmological parameters. A useful method to obtain cosmological constraints, without performing
a complete perturbative analysis, is to reduce the full likelihood information to a few parameters: the acoustic scale, $l_{A}$, the shift parameter, $R$, and the decoupling redshift, $z_{*}$ \cite{Planck:2015XIV}.
Although these distance posteriors are almost independent on the dark energy model used \cite{Wang:2012}, they
are sensitive (mainly $R$) to the growth of the perturbations. Since we have not considered
a self-consistent brane perturbation theory, these compressed CMB information could
lead to a ($\Lambda$CDM) biased brane tension constraint \cite{Alam:2016wpf,Planck:2015XIV}.
Nevertheless, as a first approach, we constrain the braneworld model parameters using the  $R$, $l_{A}$, and $z_{*}$ 
values for a flat $w$-CDM obtained from Planck measurements ($R=1.7492\pm0.0049$, $l_{A}=301.787\pm0.089$, 
$z_{*}=1089.99\pm0.29$ \cite{Neveu:2016}). 
Planck also provide the following inverse covariance matrix, $\mbox{Cov}^{-1}_{Pl}$, of these quantities
\begin{equation}
 \mbox{Cov}^{-1}_{Pl}= \left(
\begin{array}{ccc}
162.48 & -1529.4 & 2.0688 \\
 -1529.4 & 207232 & -2866.8 \\
2.0688 & -2866.8 & 53.572 \\
\end{array}
\right).
\end{equation}
The merit function for the Planck data is constructed as
\begin{equation}\label{cmbchi}
 \chi^2_{Pl} = X^T\,\mbox{Cov}_{Pl}^{-1}\,X,
\end{equation}
where
\begin{equation}
 X =\left(
 \begin{array}{c}
 l_A^{th} - l_A \\
 R^{th} -  R\\
 z_{*}^{th} - z_{*}
\end{array}\right),
\end{equation}
and the superscript $th$ refers to the theoretical estimations. The shift parameter is defined as \cite{Bond:1997wr}
\begin{equation}
R = \sqrt{\Omega_{m}H_{0}^2} r(z_{*}),
\label{eq:R}
\end{equation}
where $z_*$ can be estimated using \cite{Hu:1995en},
\begin{equation}
z_* = 1048[1+0.00124(\Omega_b h^2)^{-0.738}]
[1+g_1(\Omega_{m}h^2)^{g_2}],
\end{equation}
and
\begin{equation}
g_1 = \frac{0.0783(\Omega_b h^2)^{-0.238}}{1+39.5(\Omega_b h^2)^{0.763}},\qquad
g_2 = \frac{0.560}{1+21.1(\Omega_b h^2)^{1.81}}.
\end{equation}
The acoustic scale is defined as
\begin{equation}
l_A = \frac{\pi r(z_*)}{r_s(z_*)},
\label{eq:lA}
\end{equation}
where $r$ is the comoving distance from the observer to redshift $z$ given by
\begin{equation}
r(z)=H_0^{-1}\int_0^z \frac{dz'}{E(z')}.
\label{eq:rz}
\end{equation}

\section{Results} \label{Res}

\begin{figure*}[h]
   \centering
       \includegraphics[width=1.2\textwidth]{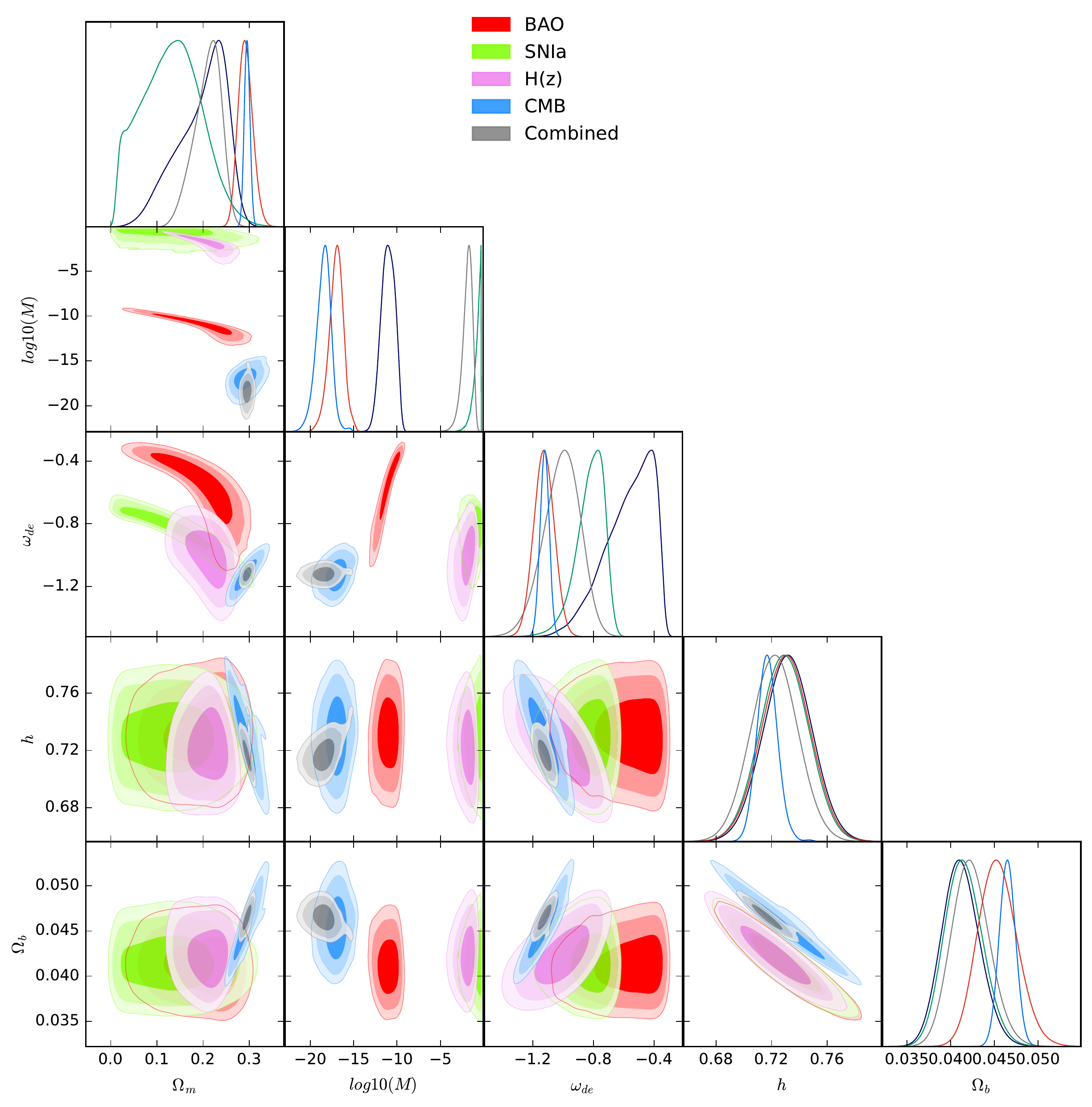} 
   \caption{1D marginalized posterior distributions and 2D contours at $68\%$, $95\%$, and $99\%$ 
   confidence levels for the $\Omega_{m}$, $\log10(\mathcal{M})$, $w_{de}$, $h$, $\Omega_{b}$ parameters using
   $H(z)$, SNIa, BAO, CMB and combined data.}
   \label{fig:contours1}               
\end{figure*}

In all our analysis, a total of $3500$ steps with $500$ walkers are generated and it takes $500$ (burn-in) steps to stabilize our estimations. Table \ref{tab:results}
summarizes the constraints and Figure \ref{fig:contours1} shows
the 1D marginalized posterior distributions and 2D contours at $68\%$, $95\%$, and $99\%$ 
confidence levels for each and combined data set. When the Hubble measurements are used, 
a lower chi-square is obtained indicating an overfitting of the data. 
The value
of $w_{de}$ is consistent with the CC and $\Omega_{m}$ is lower than the standard prediction.
Additionally, the bound of $\mathcal{M}$ suggests slightly corrections to standard dynamics of the Universe.
As has been mentioned in \S \ref{subsec:hz}, the $H(z)$ measurements rely on low redshift information and give
low significance brane constraints.
We note that the SNIa data provide a good fit of the model parameters, the dark energy EoS is consistent with the CC at $2\sigma$ level
and the limit for $\mathcal{M}$ is marginally consistent with modifications to 
gravity by an extra dimension. Nevertheless, the SNIa data
give a low value for the matter content, which is not  compatible with the one provided by the $\Lambda$CDM model.
Therefore, the SNIa data are not able to fit a precise value of $\Omega_{m}$ in brane models \cite{BraneSN:2006}
and hence the $\mathcal{M}$ bound is not statistically significant.
The BAO probe gives a lower $\Omega_{m}$ bound, a $w_{de}$ value consistent
with quintessence, and a negligible $\mathcal{M}$ value which indicates that the cosmological dynamics
of the Universe is not affected by an extra dimension. 
As we mentioned in \S \ref{subsec:bao}, the brane constraints derived from BAO data could be biased because
we assume the validity of general relativity and thus, as first approximation no brane corrections were take into account.
The CMB distance posteriors provide strong constraints on $\Omega_{m}$
and $w_{de}$, which are consistent with the $\Lambda$CDM paradigm, and no extra dimension is preferred. 
However, the compressed CMB data used here may not be appropriate to our braneworld model and these bounds could be biased. 
A joint analysis (H(z)+SNIa+BAO+CMB distance posteriors) 
estimates $\Omega_{m}$, $w_{de}$ values consistent with those of 
the standard model and the $\mathcal{M}$ bound suggests a canonical background dynamics.

\begin{figure*}[h]
   \centering
        \label{fig:contour1}         
        \includegraphics[width=0.6\textwidth]{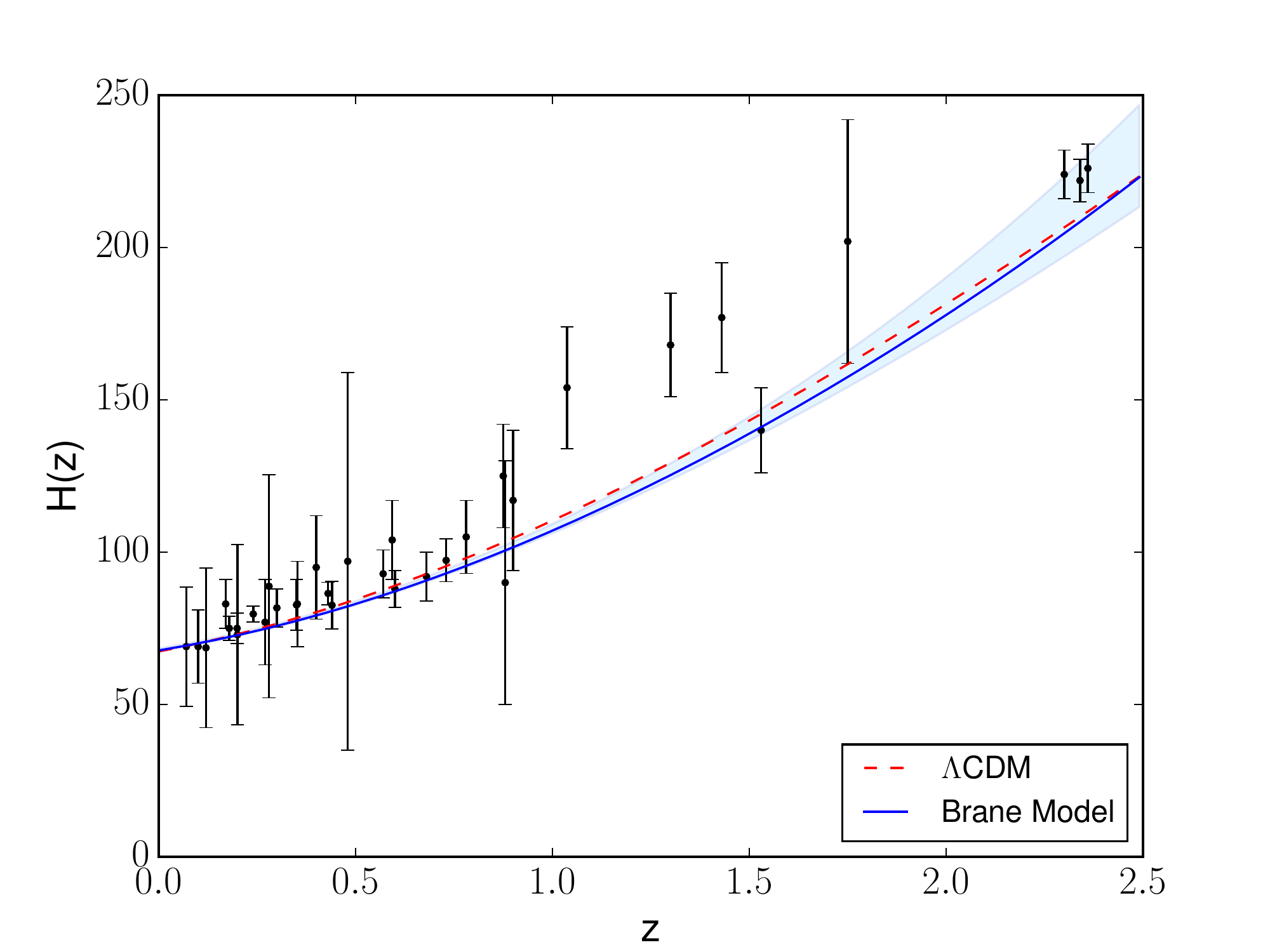}           
       \label{fig:contour2}         
        \includegraphics[width=0.6\textwidth]{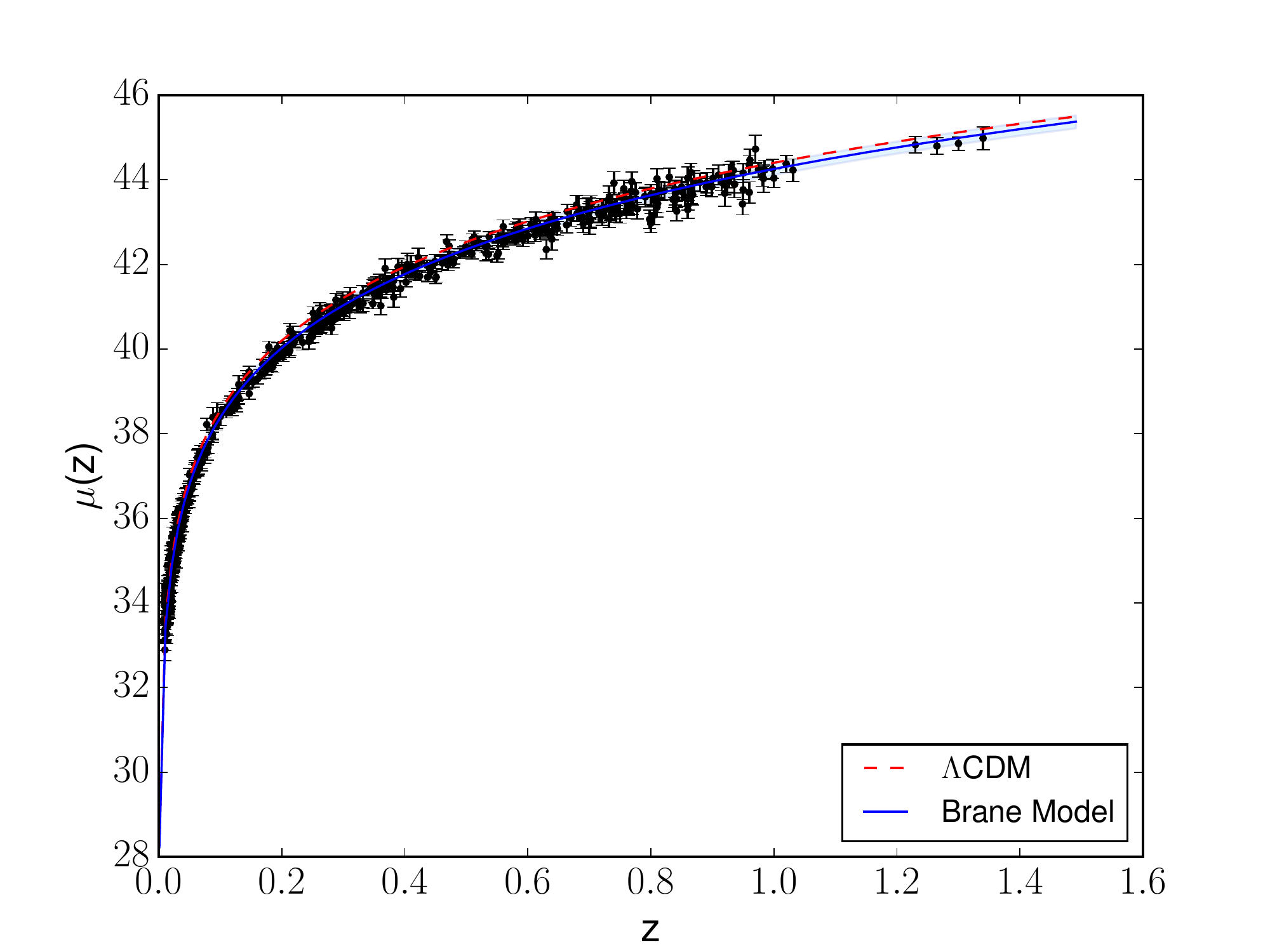} 
   \caption{Best fits to $H(z)$ and SNIa data sets for $\Lambda$CDM model (red dashed-line) and brane model (blue solid-line). 
    The blue band is the $68\%$ of confidence region of brane model corresponding to the uncertainties of $\mathcal{M}$ presented in Table \ref{tab:results}, 
    leaving the rest of the parameters fixed to their best values.}
   \label{fig:LCDMvsBrane}                
\end{figure*}

Our results show that all data sets provide $w_{de}$ bounds compatible with the CC at the $3\sigma$ confidence
level. In order to determine the deviation of the brane model from the $\Lambda$CDM
we estimate the standard model parameters taking into account a flat prior for $\Omega_m$, and Gaussian priors for $h$ and $\Omega_b$ 
as shown in Table \ref{tab:priors}, and fixed $w_{de}=-1$. We provide the $\Lambda$CDM constraints in Table \ref{tab:results}  and  
Figure \ref{fig:LCDMvsBrane} shows the fit to the $H(z)$ and SNIa data
for the brane model (solid lines) and $\Lambda$CDM (dashed lines).
Notice that the former model is consistent with the standard one within 1$\sigma$ of confidence region for both data sets.

\begin{figure*}[h]
   \centering
        \label{fig:contour1}         
        \includegraphics[width=0.6\textwidth]{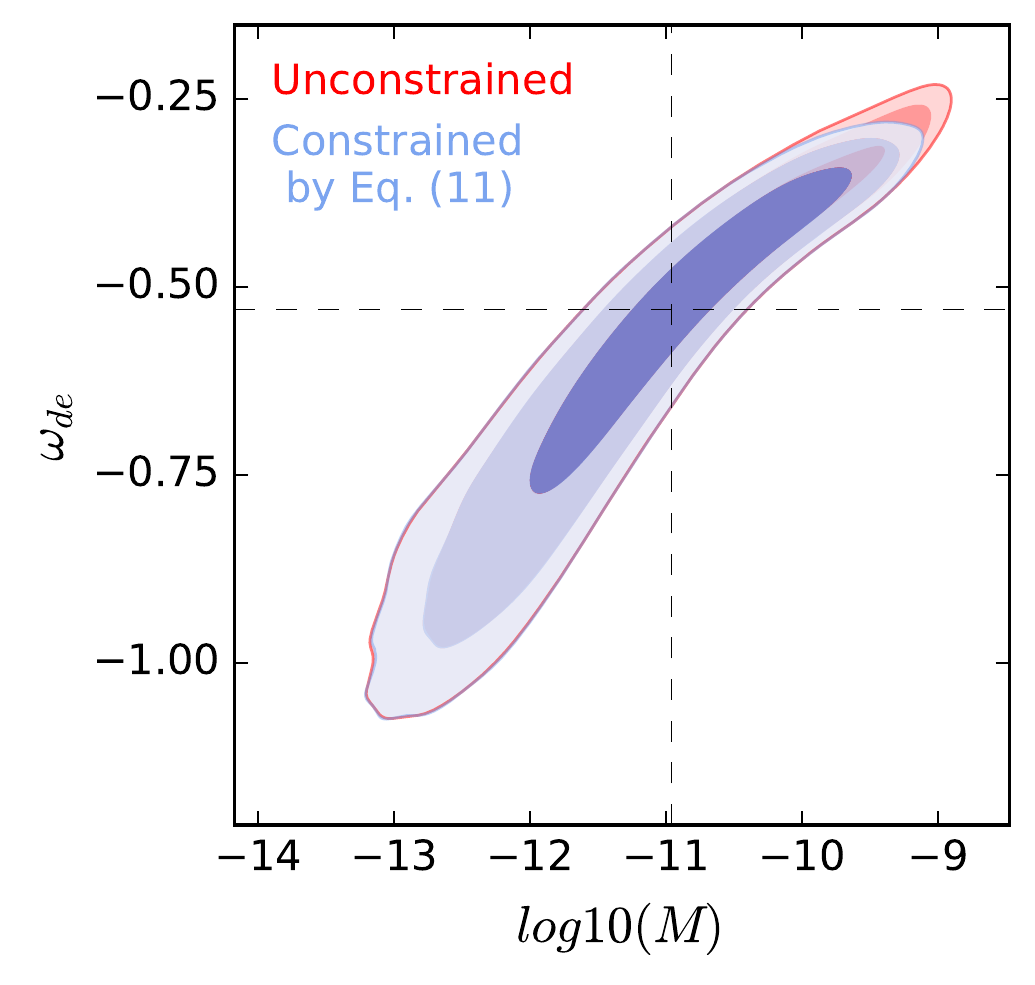}           
       \label{fig:contour2}         
        \includegraphics[width=0.6\textwidth]{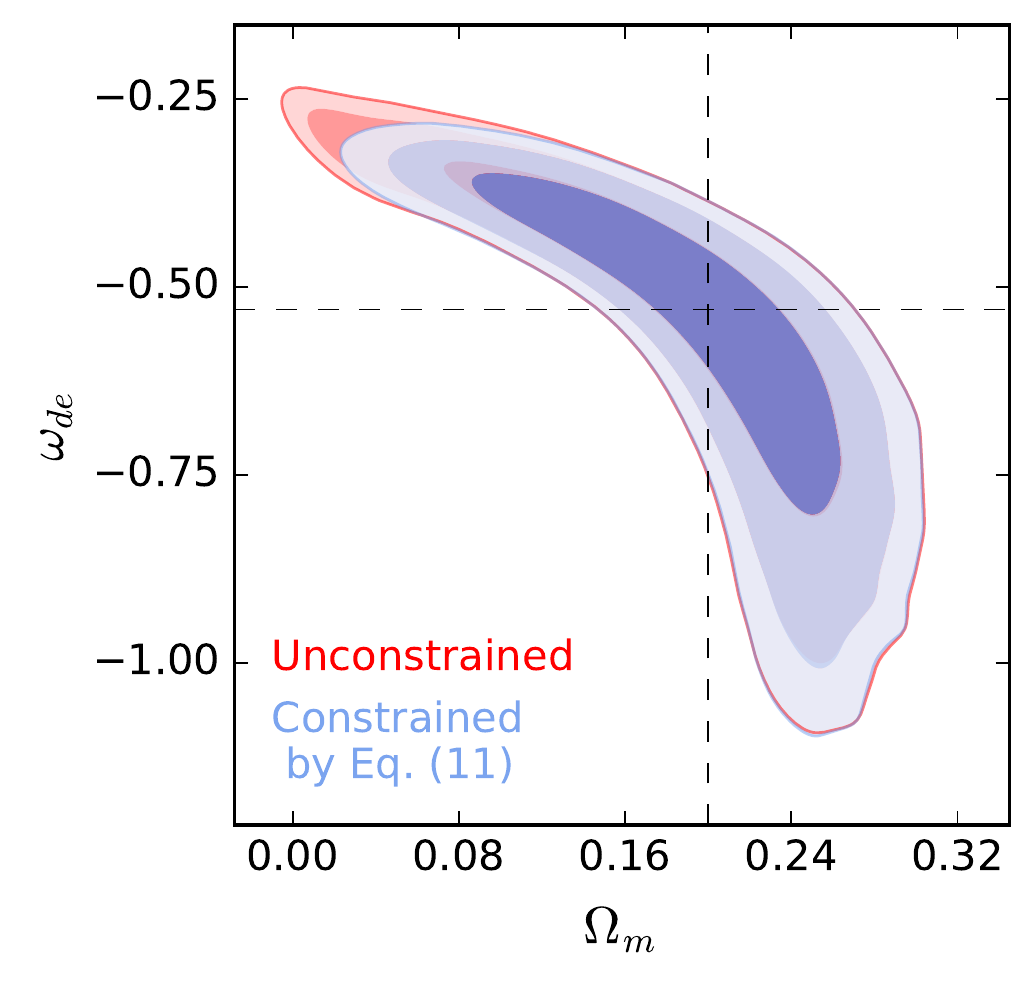} 
   \caption{
    Comparison of contours (un)constrained by Eq. \eqref{eq:de_constraint} for BAO data: $w_{de}$ vs $\log_{10}(\mathcal{M})$ (top) and $w_{de}$ vs $\Omega_{m}$ (bottom). Notice that 
Eq. \eqref{eq:de_constraint} must be satisfies to obtain a late cosmic acceleration. 
    In both figures, the unconstrained best fit is represented with dashed lines.}
   \label{fig:contours3y5}                
\end{figure*}

\begin{figure*}[h]
   \centering
        \label{fig:qz_hz}         
        \includegraphics[width=0.4\textwidth]{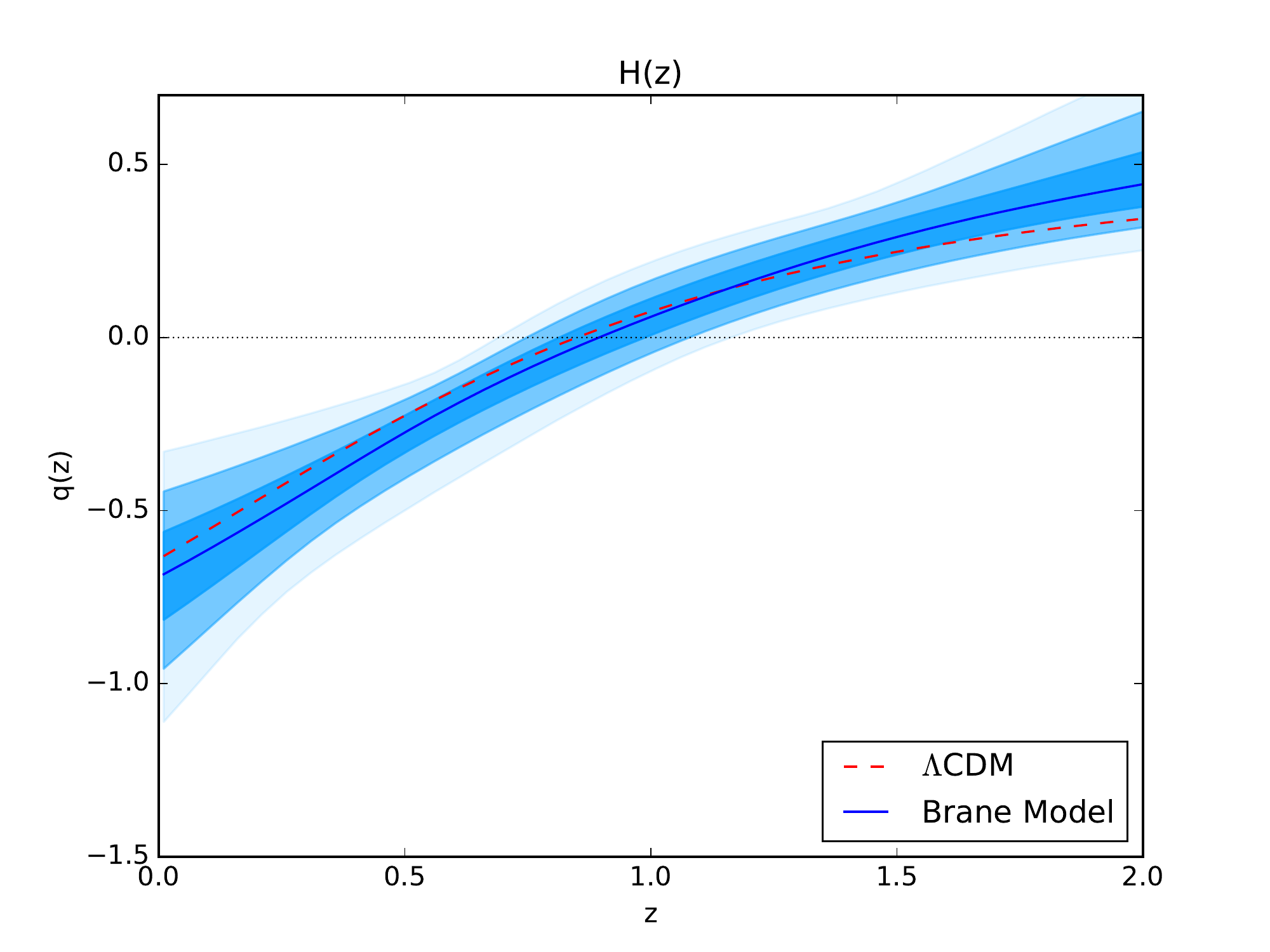}            
       \label{fig:qz_sn}         
        \includegraphics[width=0.4\textwidth]{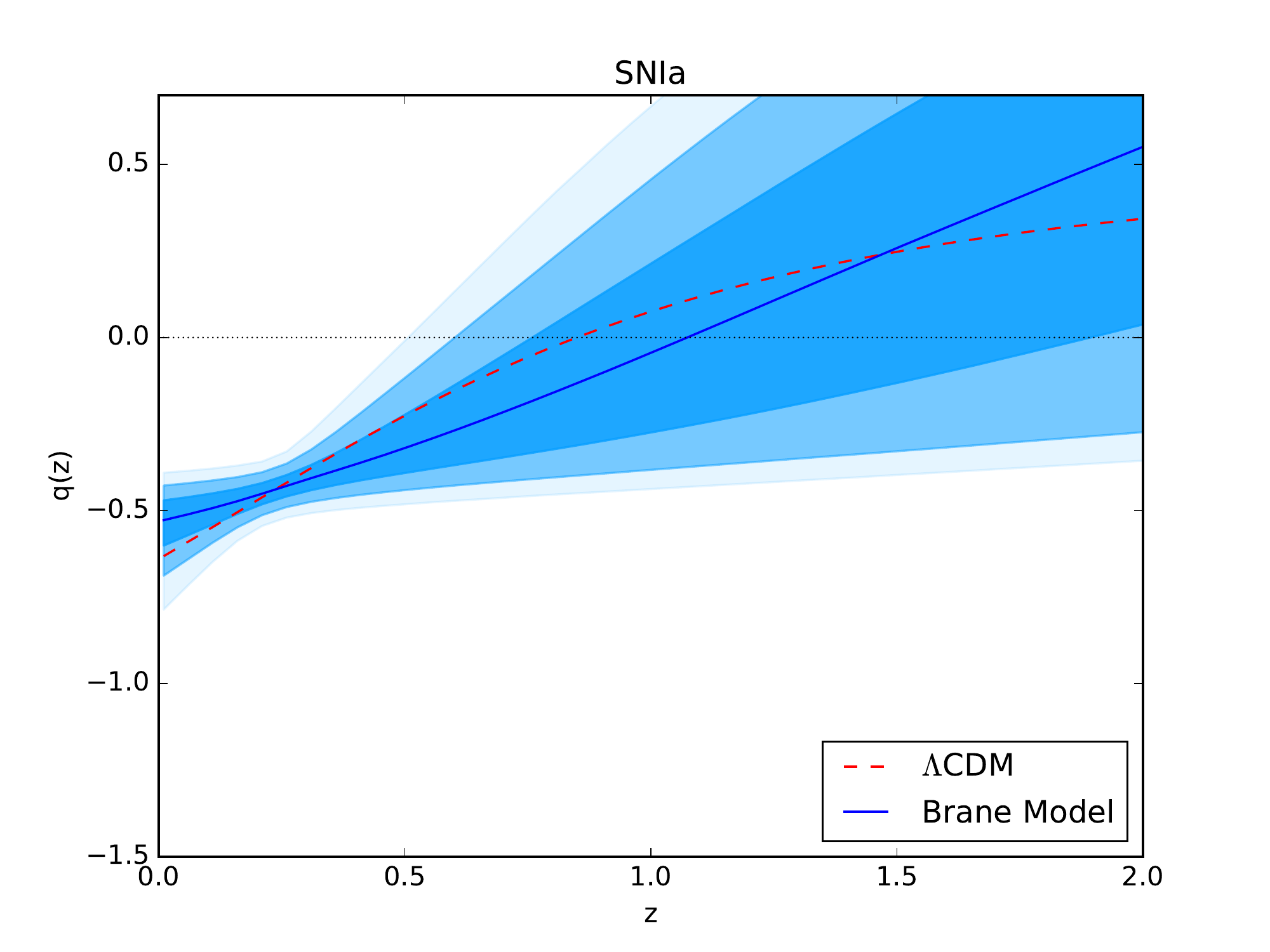}  \\
        \label{fig:qz_bao}         
        \includegraphics[width=0.4\textwidth]{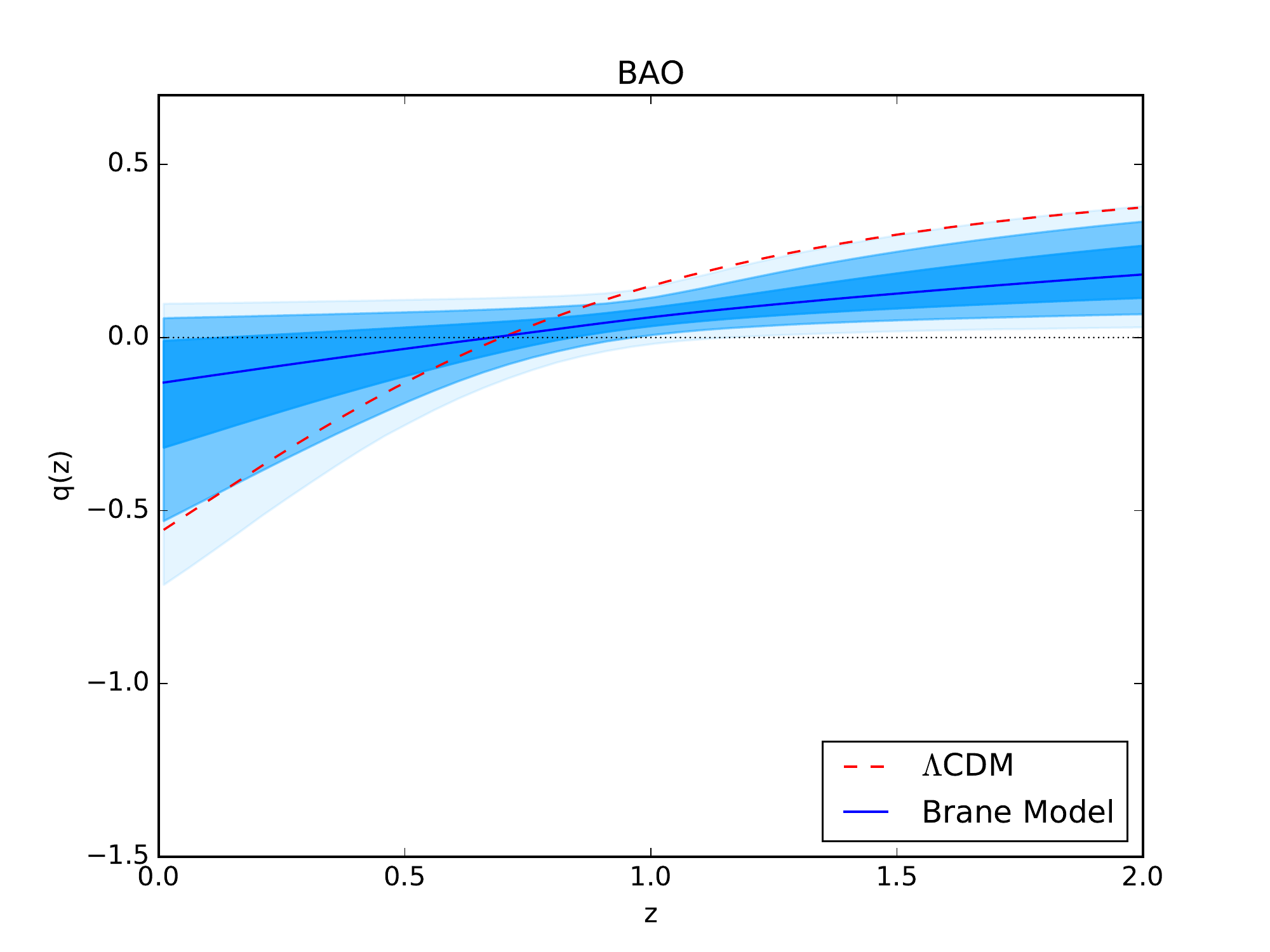}            
       \label{fig:qz_cmb}         
        \includegraphics[width=0.4\textwidth]{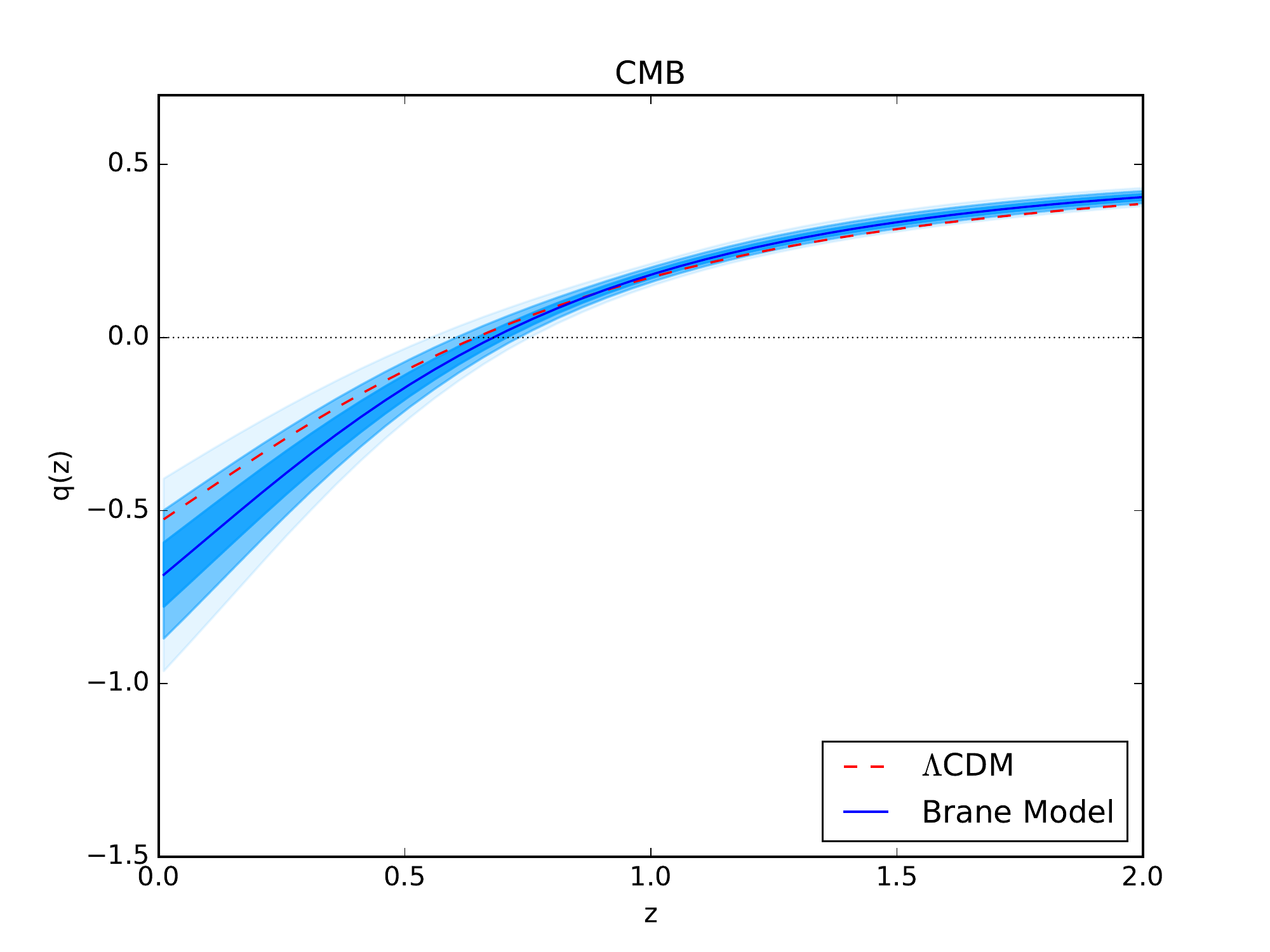} 
   \caption{Reconstruction of the deceleration parameter $q(z)$ using the constraints
   from H(z), SNIa, BAO and CMB data for two cases: the brane model (solid lines) and $\Lambda$CDM (dashed lines). The uncertainty bands correspond to $1$, $2$ and $3$ sigmas of the brane model and the dotted horizontal line represents the null acceleration.}
   \label{fig:q_plot}                
\end{figure*}

In addition, it is important to investigate whether the brane model limits satisfy the DE EoS constraint given by the Eq. (\ref{eq:de_constraint}) to guarantee a late-time cosmic acceleration.
For each cosmological data, we plot the confidence contours for the parameters
with and without this constraint (Fig. \ref{fig:contours3y5}), finding differences only for the BAO limits. Notice that our best fits satisfy the
Eq. (\ref{eq:de_constraint}) implying a late-time cosmic acceleration.
Furthermore, we reconstruct the cosmological evolution of the deceleration parameter (Eq. \eqref{eq:qz}). 
Figure \ref{fig:q_plot} shows $q(z)$ vs. redshift, using the different cosmological constraints, in two cases: the brane model
(solid lines) and $\Lambda$CDM model (dashed lines). 
When the $H(z)$ brane constraints are used to reconstruct the $q(z)$ parameter, we found that it is
slightly deviated but consistent at the $2\sigma$ confidence level with the standard dynamics.
A similar result is found using the SNIa brane constraints. 
When the BAO brane limits are included, an important tension with the $\Lambda$CDM prediction is found.
Nevertheless, the $q(z)$ evolution is consistent with the standard one at the $3\sigma$ confidence level. 
Using the CMB brane constraints, the $q(z)$ evolution shows a slightly deviation from of the standard model, but it
is consistent at the $2\sigma$ confidence level.

\begin{figure*}[h]
   \centering
     \includegraphics[width=0.7\textwidth]{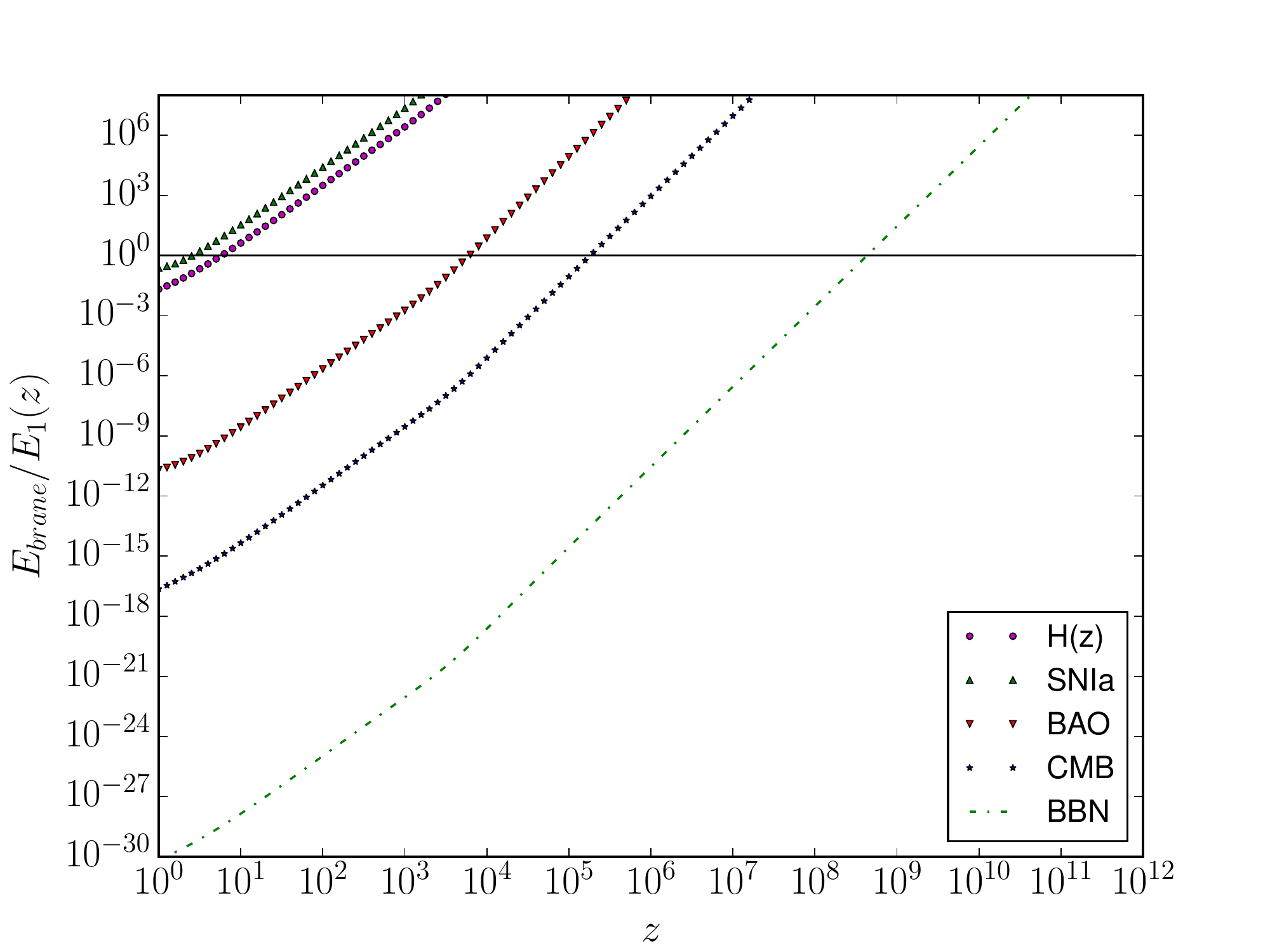} 
   \caption{Ratio of the corrective term in Eq. (\ref{eq:H}) by the brane dynamics and the first canonical
   term in the Friedmann equation vs. redshift using the different cosmological constraints shown in Table \ref{tab:recompilation}. Note that when the ratio is greater than $1$,
   the effect due to the brane dynamics becomes dominant.}
   \label{fig:ratio}                
\end{figure*}

We found different upper limits on the $\mathcal{M}$ constraints, which is directly
related to the Friedmann equation modifications and hence, to the Universe dynamics. While the low-redshift data ($H(z)$ and SNIa) suggest slightly
evidence of gravity modifications, the high-redshift data (BAO and CMB distance constraints) rule
out an extra dimension in the Universe. 
Fig. \ref{fig:ratio} shows the ratio of the corrective term in Eq. (\ref{eq:H}) by the brane dynamics and the first canonical term in the Friedman equation vs. redshift using the different cosmological constraints. 
When the ratio is greater than one, the effect due to the brane dynamics becomes dominant. 
This confirms that the brane effect become important at different energy scales (redshifts)
for each observational data. Furthermore, its effects are larger at higher redshifts (see Eq. \ref{eq:H}).
Thus, the $\mathcal{M}$ bounds provided by low redshift probes are less significant than the high redshift ones.

There is a huge discrepancy in the brane tension bounds derived from cosmological tests and those obtained 
from BBN \cite{Maartens:2003tw}, stellar dynamics \cite{yo2,gm,Garcia-Aspeitia:2015mwa,Linares:2015fsa} and TT experiments \cite{Gergely}. 
The brane tension lies on $\lambda_{\mathrm{BBN}}>1\,\rm {MeV}^4$ in the first one and $\lambda_{\mathrm{TT}}\gtrsim138\,\rm TeV^4$ in the latter one
(for a compilation see Table \ref{tab:recompilation}). 
Our most restricted bound for the brane tension (Table \ref{tab:results})
is $\lambda_{\mathrm{CMB}}>4.05\times10^{4}h^{2}\mathrm{eV^{4}}$ 
($\lambda_{joint}> 6.42\times10^{5}h^{2}\mathrm{eV^{4}}$ for the joint analysis),
while it should be $\lambda\gg \lambda_{\mathrm{joint}}.$ 
 In spite of this, we provide new constraints (in the low energy regime) on the brane tension which are complementary to those shown in Table \ref{tab:recompilation}.

Finally, a further analysis of the brane perturbations would give information about the viability of the model. In this vein, Ref. \cite{Koyama:2003be} explore the consequences of a brane simple model without dark radiation on the CMB spectrum. The authors show that at large scales 
the temperature anisotropy caused by Sachs-Wolfe effect is the same as the canonical one. They also claim that 
at very small scales the effects of branes are negligible. Nevertheless, on scales up to the first CMB acoustic peak,  
the brane terms considerably modify its amplitude and position. This implies a change in the CMB distance posteriors and, thus, in the brane constraints that we have obtained. To asses the impact of the perturbation on the brane constraints, a full CMB analysis should be carried out, which is beyond of the scope of this article.

\begin{table}
\tbl{Best fit values and their uncertainties for brane ($\Omega_{m}$, $w_{de}$, $\log(\mathcal{M})$) and $\Lambda$CDM ($\Omega_m$) models estimated from 
H(z), SNIa, BAO, CMB distance constraints and a joint analysis (H(z)+SNIa+BAO+CMB distance constraints). 
The brane tension is established by a lower limit at $99\%$ of CL and is calculated (see Eq. \eqref{const}) assuming 
a critical density $\rho_{crit}=8.10h^2\times10^{-11}\rm eV^4$.}
{\begin{tabular}{@{}ccccccc@{}} \toprule
Model&$\chi^{2}$ & $h$ & $\Omega_{m}$&$w_{de}$&$\log(\mathcal{M})$& $\lambda$($h^2\rm eV^4$)\\ \colrule
\multicolumn{7}{c}{$H(z)$}\\ 
Brane & $18.19$& $0.72^{+0.01}_{-0.01}$&$0.21^{+0.02}_{-0.03}$& $-1.00^{+0.11}_{-0.12}$&$<-0.88$& $>3.07\times10^{-10}$\\
$\Lambda$CDM & $17.15$& $0.71^{+0.01}_{-0.01}$ & $0.24^{+0.01}_{-0.01}$&$-1.0$& \textemdash\textemdash\textemdash& \textemdash\textemdash\textemdash\\ \botrule
\multicolumn{7}{c}{SNIa}\\ 
Brane & $574.73$& $0.72^{+0.01}_{-0.01}$& $0.13^{+0.06}_{-0.07}$& $-0.81^{+0.07}_{-0.10}$&$<-0.31$ &$>8.27\times10^{-11}$ \\
$\Lambda$CDM & $576.12$& $0.72^{+0.01}_{-0.01}$ & $0.24^{+0.01}_{-0.01}$&$-1.0$& \textemdash\textemdash\textemdash& \textemdash\textemdash\textemdash \\ \botrule
\multicolumn{7}{c}{BAO}\\ 
Brane & $5.46$ & $0.73^{+0.01}_{-0.01}$& $0.20^{+0.04}_{-0.07}$&$-0.53^{+0.13}_{-0.19}$&$<-9.52$&$>0.13$\\
$\Lambda$CDM & $13.95$& $0.66^{+0.01}_{-0.01}$ & $0.29^{+0.02}_{-0.02}$&$-1.0$& \textemdash\textemdash\textemdash& \textemdash\textemdash\textemdash \\ \botrule
\multicolumn{7}{c}{CMB distance constraints}\\ 
Brane & $10.87$&$0.73^{+0.01}_{-0.01}$& $0.29^{+0.01}_{-0.01}$&$-1.12^{+0.06}_{-0.06}$&$<-15.0$&$>4.05\times10^4$\\
$\Lambda$CDM & $0.94$& $0.68^{+0.005}_{-0.005}$ & $0.31^{+0.008}_{-0.008}$&$-1.0$&\textemdash\textemdash\textemdash& \textemdash\textemdash\textemdash  \\ \botrule
\multicolumn{7}{c}{Joint analysis }\\ 
Brane & $636.70$&$0.71^{+0.01}_{-0.01}$& $0.30^{+0.01}_{-0.01}$&$-1.12^{+0.03}_{-0.03}$&$<-16.2$&$>6.42\times10^5$ \\
$\Lambda$CDM & $640.79$& $0.68^{+0.004}_{-0.004}$ & $0.30^{+0.005}_{-0.005}$&$-1.0$&\textemdash\textemdash\textemdash&  \\\botrule
\end{tabular} \label{tab:results}}
\end{table}

\begin{table}
\tbl{Summary of the brane tension constraints derived from several experiments and observations. 
The last value corresponds to our best constraint using the joint analysis from Table \ref{tab:results}, 
but in comparison it remains the weakest.}
{\begin{tabular}{@{}cc@{}} \toprule
Experiment/Observation & Cut-off ($\rm eV^4$)\\ \colrule
Table-Top& $138.59\times10^{48}$,\;\; \cite{Gergely}\\
Astrophysical& $5\times10^{32}$, \;\; \cite{yo2,gm,Garcia-Aspeitia:2015mwa,Linares:2015fsa}\\
BBN & $10^{24}$, \;\; \cite{MaartensR}\\
Joint analysis & $6.42h^2\times10^5$\\
\botrule
\end{tabular} \label{tab:recompilation}}
\end{table}

%
%

\section{Conclusions and Remarks} \label{CR}

Brane theory is an interesting paradigm with the potential to solve many fundamental problems in Particle Physics 
and Cosmology, being an important candidate to extend GR. In this paper, we explored the consequences 
in the cosmic acceleration by considering a generic dark energy in a Randall-Sundrum braneworld scenario. 
The modified Friedmann equations governing the dynamics of the Universe were derived to investigate
whether the current cosmological observations suggest such gravity corrections.
We put constraints on the dark matter density parameter, dark energy EoS, and the brane tension using
the latest observational data (H(z), SNIa, BAO, and CMB distance constraints
from Planck data release 2015). In particular, we provide brane tension lower limits 
in the low-energy regime (low redshift), complementary to those obtained in the high-energy regime.

We found that all data sets provide dark energy equation of state ($w_{de}$) compatible with the CC at the $3\sigma$ confidence level. 
Different bounds in the brane tension were estimated using the cosmological data.  
While the high-redshift data (BAO and CMB) prefer no gravity modifications, 
the low-redshift data available ($H(z)$ and SNIa) slightly suggest that there is an extra dimension. 
However, as the brane effect is more important at higher redshifts, the bounds obtained from low redshift probes are less significant than the ones obtained from those at high redshift. 
Furthermore, a joint analysis of the cosmological data provides a 
tight constraint for the brane tension but the huge discrepancy with complementary observations/experiments persists. It is worth to note that a self-consistent brane perturbation analysis for this model on high-redshift data is needed in order to asses its effect on our constraints. 

We reconstructed the deceleration parameter using the best fit for each dataset
and found that the dark energy component drives to a late-time cosmic acceleration 
independently of gravity modifications by an extra dimension. 

Finally, an appropriate extension of the modified FLRW equations for braneworld models is needed. For instance, by considering the crossed terms which take into account coupling between the different Universe 
components or the consideration of a variable brane tension.
However, this work is out of our present scope.

\section*{Acknowledgments}
We would like to thank the referee for thoughtful comments which helped to improve the manuscript. 
MAG-A acknowledges support from SNI-M\'exico and CONACyT research fellow. 
J.M. acknowledges support from CONICYT/FONDECYT 3160674. A.H. acknowledges CONACyT postdoctoral fellow; Instituto Avanzado de Cosmolog\'ia (IAC) collaborations.
V.M. acknowledges support from Centro de Astrof\'{\i}sica de Valpara\'{\i}so.

\bibliographystyle{ws-ijmpd}
\bibliography{librero1}

\end{document}